\documentclass[a4paper,fleqn,usenatbib,useAMS]{mnras}

\usepackage[T1]{fontenc}
\usepackage{ae,aecompl}

\usepackage{graphicx}	
\usepackage{amsmath}	
\usepackage{amssymb}	

\usepackage{times}
\usepackage{epstopdf}
\usepackage{graphicx}
\usepackage{txfonts}
\usepackage{natbib}
\usepackage{array}
\title[The hotspots of Cygnus A]{LOFAR imaging of Cygnus A -- Direct detection of a turnover in the hotspot radio spectra}

\title[The hotspots of Cygnus A]{LOFAR imaging of Cygnus A -- Direct detection of a turnover in the hotspot radio spectra}
\author[J. P. McKean et al.]{J.~P.~McKean$^{1,2}$, L.~E.~H.~Godfrey$^{1}$, S.~Vegetti$^{3}$, M.~W.~Wise$^{1,4}$, R.~Morganti$^{1,2}$, 
\newauthor M.~J.~Hardcastle$^{5}$, D.~Rafferty$^{6}$,  J.~Anderson$^{7}$,  I.~M.~Avruch$^{2,8}$, R.~Beck$^{9}$, M.~E.~Bell$^{10}$,
\newauthor I.~van Bemmel$^{11}$, M.~J.~Bentum$^{1,12}$, G.~Bernardi$^{13,14}$, P.~Best$^{15}$, R.~Blaauw$^{1}$,
\newauthor A.~Bonafede$^{6}$, F.~Breitling$^{16}$, J.~W.~Broderick$^{17}$, M.~Br\"uggen$^{6}$, L.~Cerrigone$^{1}$,
\newauthor B.~Ciardi$^{3}$, F.~de Gasperin$^{18}$, A.~Deller$^{1}$, S.~Duscha$^{1}$, D.~Engels$^{19}$, H.~Falcke$^{1,20}$,
\newauthor R.~A.~Fallows$^{1}$, W.~Frieswijk$^{1}$, M.~A.~Garrett$^{1,18}$, J.~M.~Grie\ss{}meier$^{21,22}$,
\newauthor M.~P.~van Haarlem$^{1}$, G.~Heald$^{1,2}$, M.~Hoeft$^{23}$, A.~J. ~van der Horst$^{24}$, M.~Iacobelli$^{1}$,
\newauthor H.~Intema$^{18,25}$, E.~Juette$^{26}$, A.~Karastergiou$^{27}$, V.~I.~Kondratiev$^{1,28}$, L.~V.~E.~Koopmans$^{2}$,
\newauthor M.~Kuniyoshi$^{29}$, G.~Kuper$^{1}$, J.~van Leeuwen$^{1,4}$, P.~Maat$^{1}$, G.~Mann$^{16}$, S.~Markoff$^{4}$,
\newauthor R. McFadden$^{1}$, D.~McKay-Bukowski$^{30,31}$, D.~D.~Mulcahy$^{17}$, H.~Munk$^{1}$, A.~Nelles$^{32}$, 
\newauthor E.~Orru$^{1}$, H.~Paas$^{33}$, M.~Pandey-Pommier$^{34}$, M.~Pietka$^{27}$, R.~Pizzo$^{1}$, A.~G.~Polatidis$^{1}$,  
\newauthor W.~Reich$^{9}$, H.~J.~.A.~R\"ottgering$^{18}$, A.~ Rowlinson$^{1}$, A.~M.~M.~Scaife$^{35}$, M.~Serylak$^{36}$,
\newauthor A.~Shulevski$^{2}$, J.~Sluman$^{1}$, O.~Smirnov$^{37,14}$, M.~Steinmetz$^{16}$, A.~Stewart$^{27}$,
\newauthor J.~Swinbank$^{38}$, M.~Tagger$^{21}$, S.~Thoudam$^{20}$, M.~C.~Toribio$^{1,18}$, R.~Vermeulen$^{1}$,
\newauthor C. Vocks$^{16}$, R.~J.~van Weeren$^{13}$, O.~Wucknitz$^{9}$, S.~Yatawatta$^{1}$ and P.~Zarka$^{39}$\\
Affiliations are listed at the end of the paper}

\date{Accepted 2016 August 19. Received 2016 August 11; in original form 2016 January 27}

\pubyear{2016}

\begin{document}
\label{firstpage}
\pagerange{\pageref{firstpage}--\pageref{lastpage}}
\maketitle

\begin{abstract}
The low-frequency radio spectra of the hotspots within powerful radio galaxies can provide valuable information about the physical processes operating at the site of the jet termination. These processes are responsible for the dissipation of jet kinetic energy, particle acceleration, and magnetic-field generation. Here we report new observations of the powerful radio galaxy Cygnus A using the Low Frequency Array (LOFAR) between 109 and 183 MHz, at an angular resolution of $\sim3.5$~arcsec. The radio emission of the lobes is found to have a complex spectral index distribution, with a spectral steepening found towards the centre of the source. For the first time, a turnover in the radio spectrum of the two main hotspots of Cygnus A has been directly observed. By combining our LOFAR imaging with data from the Very Large Array at higher frequencies, we show that the very rapid turnover in the hotspot spectra cannot be explained by a low-energy cut-off in the electron energy distribution, as has been previously suggested. Thermal (free-free) absorption or synchrotron self absorption models are able to describe the low-frequency spectral shape of the hotspots, however, as with previous studies, we find that the implied model parameters are unlikely, and interpreting the spectra of the hotspots remains problematic.
\end{abstract}





\section{Introduction}
\label{intro}

In powerful radio galaxies, the active galactic nucleus generates two oppositely directed jets of plasma with a relativistic bulk velocity. These jets terminate and convert much of their bulk kinetic energy into particles and magnetic fields at regions where they impact with the ambient medium at the outer edge of the lobe. This working surface is identified with hotspots, which typically have a compact structure and a high surface brightness. The impact of the jet generates a reverse shock that propagates upstream into the jet, and is responsible for the conversion of the bulk kinetic energy into random particle energies and turbulence. The jet material flows through the hotspots and back towards the nucleus, spreading out to form the extended radio lobes (e.g. \citealt{blandford74}).

Observations of powerful radio galaxies, when combined with synchrotron spectral ageing models, show good agreement with the scenarios described above (e.g.\,\citealt{myers85,carilli91,harwood13}). However, in a small number of high power objects ($L_{\rm 178~MHz} > 10^{27}$~W~Hz$^{-1}$~sr$^{-1}$), the hotspot radio spectra are seen to flatten towards low frequencies ($<$~327~MHz; e.g.\,\citealt{leahy89}), with the spectrum becoming flatter than the canonical $S_\nu \propto \nu^{-0.5}$ spectrum predicted for diffusive shock acceleration \citep[see][and references therein]{godfrey09}. This flattening has been generally interpreted as a low-energy cut-off (LEC) in the energy distribution of the relativistic particles within the hotspots, where a spectral turnover is expected to be seen at low frequencies. However, this has never been observationally verified due to the limited spectral coverage and angular resolution of low frequency radio telescopes.

Cygnus A (3C\,405; $z =0.0561$) is the nearest and most powerful example of a classical double radio galaxy, and as such, is considered to be the archetype of this class of radio source. Cygnus A is one of the first objects in which the hotspot radio spectra were seen to flatten towards low frequencies \citep{leahy89}, but the physical cause of this spectral flattening has been a source of contention. \citet{carilli91} performed a detailed multi-frequency analysis of Cygnus A at 4.5~arcsec resolution, based on imaging with the Very Large Array (VLA) at several frequencies between 327 MHz and 22~GHz, along with 3.0~arcsec resolution Multi-Element Radio Linked Network (MERLIN) imaging at 151~MHz taken by \citet{leahy89}. At frequencies above 1~GHz, they found the spectra of the two secondary hotspots (A and D) to be well described by a broken power law of the form predicted by a continuous injection model, in which a power-law electron energy distribution is continuously injected into the hotspot, and radiative cooling acts to modify the electron distribution as it flows away from the hotspots and into the lobes. In addition, the injected particle spectrum was found to agree with the predictions from diffusive shock acceleration. 

However, below 1 GHz the hotspot spectra were observed to deviate significantly from the broken power law model, becoming flatter towards lower frequencies. A LEC model, with a cut-off Lorentz factor of $\gamma_{\rm cr}^A = 420\pm20$ and $\gamma_{\rm cr}^D = 440 \pm 20$, for hotspots A and D, respectively, was found to reproduce the observed spectra of the hotspots down to 151~MHz, with predicted magnetic field strengths that were in agreement with those expected from equipartition arguments ($B_{\rm eq} = 300~\mu$G). Furthermore, thermal (free-free) and synchrotron-self absorption models predicted unusually high particle densities ($n_e = 2$~cm$^{-3}$ for a $10^4$~K gas)  or magnetic field strengths ($B_{\rm SSA} = 3$~G), and so, were not thought to be plausible explanations for the flattening of the hotspot spectra.

There is evidence that at frequencies $\la 100$~MHz the hotspot spectra may be turning over much more steeply than is allowed by the LEC model. First, the lack of scintillation seen from Cygnus A at $\sim 80$~MHz was thought to be due to a rapid turnover in the radio spectra of the compact hotspots \citep{muxlow88}. Second, \citet{lazio06} estimated the emission from hotspots A and D at 74 MHz by using 10 arcsec-resolution imaging with the VLA. This was done by using a spectral model for the lobe and counter-lobe emission and assuming that the excess flux was due to the hotspots (here we define the lobe as being associated with the brightest radio jet, which is expected to be on the approaching side of the source). They estimated that hotspot D had a flat spectrum consistent with a LEC model, but that the spectrum of hotspot A was decreasing too rapidly towards lower frequencies to be explained by a LEC. An analysis of the lobe spectra by \citet{kassim96} between 74 and 325 MHz at 25 arcsec resolution (where the beam dilution of the hotspots would be high) found that the lobe and counter-lobes have an asymmetric spectral index, with the counter-lobe (containing hotspot D) having a flatter spectral index than the lobe (containing hotspot A). As we will show, in contrast to the results of \citet{lazio06} and \citet{kassim96}, we find the spectrum of hotspot D in the 100 to 200~MHz range to have the more extreme curvature of the two hotspots, with a flatter slope being seen in hotspot A. 

In this paper, we report new observations of the powerful radio galaxy Cygnus A at $\sim3.5$~arcsec resolution using the Low Frequency Array (LOFAR; \citealt{vanhaarlem13}) at frequencies from 109 to 183 MHz. These data show direct evidence for a turnover in the radio spectra of both hotspots for the first time, and are combined with data taken at higher frequencies to constrain models for the jet-emission and the magnetic field in the hotspot regions. We defer a detailed discussion of the lobe emission to a follow-up paper. In Section \ref{data}, we present our observations with LOFAR and the data reduction steps, and our results are presented in Section \ref{results}. We analyze the radio spectra of the hotspots in Section \ref{discuss} and present our conclusions in Section \ref{conc}.

Throughout, we follow the same standard nomenclature for the naming of the lobes and hotspots of Cygnus A (e. g. \citealt{carilli91}) and 
define the spectral index as $S_{\nu} \propto \nu^{\alpha}$. At the redshift of Cygnus A ($z =0.0561$), 1 arcsec corresponds to a projected distance of 1.089~kpc for a $\Omega_{\rm M} = 0.3$, $\Omega_{\Lambda} = 0.7$ flat-cosmology with $H_{0} = 70$~km\,s$^{-1}$~Mpc$^{-1}$.

\section{Observations \& Data Analysis}
\label{data}

\subsection{Observations}
Cygnus A was observed during commissioning time with LOFAR using the High Band Antenna (HBA) array on 2013 March 2. The data were taken using the 200 MHz sampling clock (2nd Nyquist zone), covering a bandwidth of 39 MHz between 102 and 197 MHz. A total of 200 sub-bands were produced, each with a bandwidth of 195 kHz that were divided into 64 spectral channels. The individual sub-bands were spread out in a non-contiguous fashion, so as to cover a broader frequency range that would better constrain any spectral turnover. However, the sub-bands were grouped together in contiguous chunks of 1.95 MHz bandwidth (10 sub-bands), with each of these chunks separated by several thousand kHz. The total integration time was 6 h with a visibility averaging time of 3 s. The array consisted of 23 core stations and 13 remote stations, which together provided a well sampled {\it uv-}plane, with baseline lengths between 70 m to 84 km.

\subsection{Absolute flux-density scale}

As there is currently no well-defined frequency dependent flux-density calibration for LOFAR, the absolute flux-density of the dataset was calculated using the known spectral energy distribution of Cygnus A and bootstrapping this to the LOFAR data; this is the same technique used for other bright sources that have been observed with LOFAR (e.g. M82; \citealt{degasperin12}).  

\begin{figure}
\centering
\includegraphics[width=9.2cm]{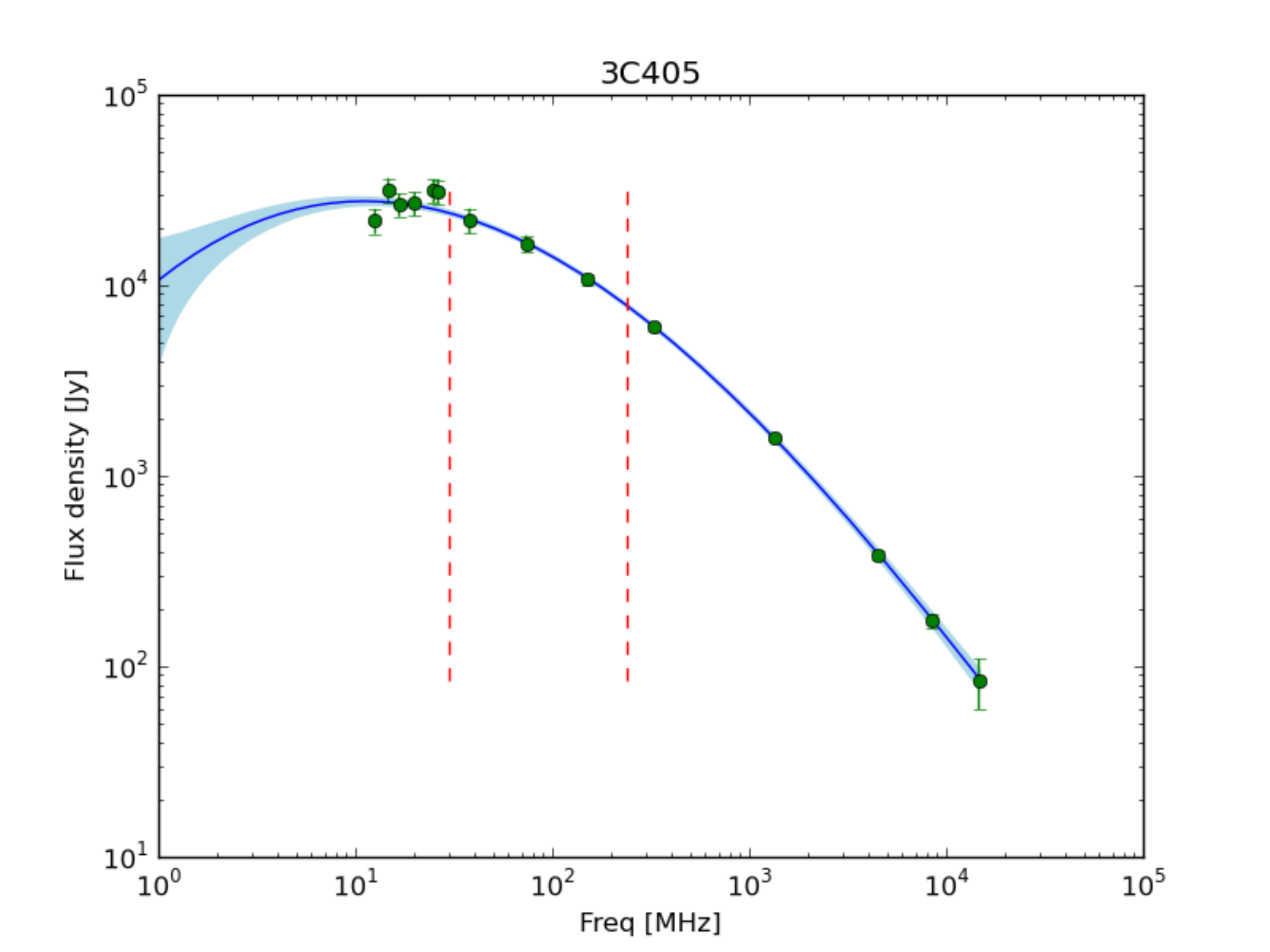}
\caption{The fitted spectral energy distribution of Cygnus A (3C\,405). The data points (green) have been taken from \citet{baars77} and \citet{carilli91}, and the best fit polynomial function (3rd order; dark blue) has been determined using the Bayesian fitting method of \citet{scaife12}. The error region of the spectrum (light blue) increases substantially below 30 MHz due to the scatter and sampling of the low frequency spectrum. However, the part of the spectrum that is covered by LOFAR (30 to 250 MHz; vertical red lines) is well sampled and the uncertainty is about 2 per cent at 150 MHz.}
\label{fig:sed}
\end{figure}

We follow the Bayesian fitting method described by \citet{scaife12} to determine the integrated radio spectrum of Cygnus A, based on previous measurements at several frequencies between 12.6~MHz and 14.6509~GHz. We find that a 3rd order polynomial of the form,
\begin{equation}
S[{\rm Jy}] = A_0 \prod^{N}_{i=1-3}10^{A_i\log^i_{10}(\nu/[{\rm 150~MHz}])},
\end{equation}
with co-efficients $A_0 = 10690\pm230$, $A_1 = -0.67\pm0.19$, $A_2 = -0.240\pm0.017$ and $A_3 = 0.021\pm0.014$, provides the best fit to the observed spectral energy distribution (see Fig.~\ref{fig:sed}). This model for the integrated flux-density of Cygnus A was used to flux-calibrate the LOFAR dataset during the data reduction stage. 

\subsection{Imaging and data analysis}

The data were analyzed using the standard LOFAR reduction packages. The radio frequency interference (RFI) were first flagged in frequency and time using the {\sc aoflagger} \citep{offringa10,offringa13}. As Cygnus A dominates the visibilities, no wide-field mapping was required. Therefore, the 64 frequency channels from each sub-band were averaged to a single frequency channel and to a visibility integration time of 10 s. The calibration was first performed manually for a select number of sub-bands that spanned the observed frequency range to establish a robust starting model for the source surface brightness distribution and its frequency dependence. As Cygnus A has complex structure on several angular-scales that was expected to be frequency dependent, the imaging of the data was carried out using the Multi-Scale--Multi-Frequency-Synthesis (MS-MFS) clean algorithm \citep{rau11} that is part of {\sc casa} (Common Astronomy Software Applications). We found that using 5 spatial scales of width 0 (point spread function),  5, 10, 15 and 20 arcsec, and 2 Taylor terms to describe the frequency dependence provided the best results. Using a larger number of sky terms was found to increase the map noise due to the differing spectral properties of the lobes and hotspots (see below).

\begin{figure*}
\centering
\includegraphics[width=15cm]{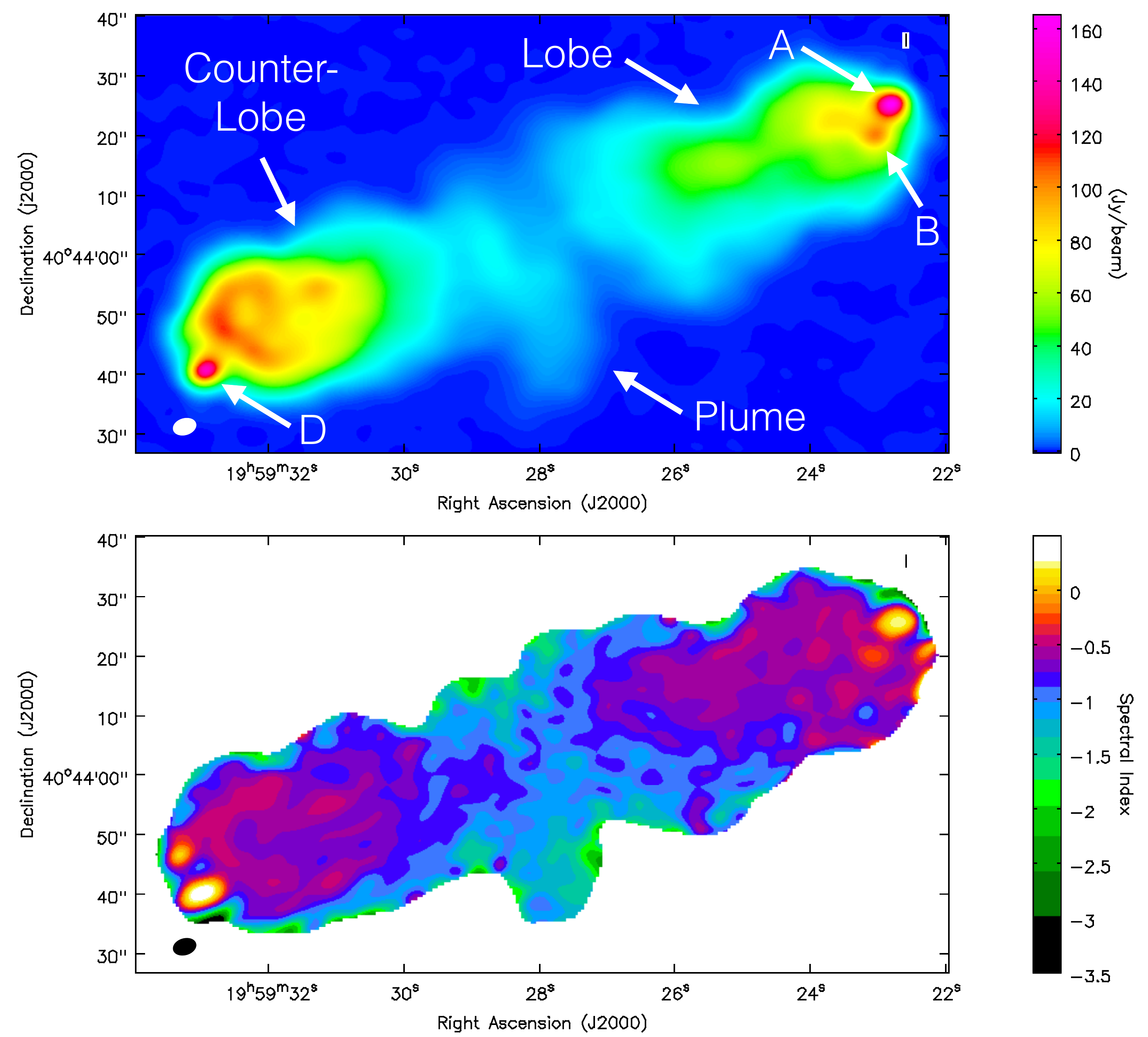}
\caption{(Upper) The total intensity map of Cygnus A between 109 and 183 MHz (a central frequency of 146~MHz). This map was made with the MS-MFS clean algorithm, using 5 spatial scales of width 0 (point spread function),  5, 10, 15 and 20 arcsec, and 2 Taylor terms to describe the frequency dependence. In order to maximize the spatial resolution of the observations and investigate the spectral properties of the hotspots, the {\it uv}-data were superuniform weighted during imaging. The resulting rms map noise is 43 mJy~beam$^{-1}$ and the FWHM beam size is $3.8\times2.7$~arcsec at a position angle of $-73$~deg east of north, shown as the filled ellipse in the bottom left corner. The dynamic range of this image is 3800. (Lower) The resulting spectral index map between 109 and 183 MHz, showing an inverted radio spectrum for the two brightest hotspots, and the expected frequency dependent structure in the lobe, counter-lobe and plume emission.}
\label{fig:maps}
\end{figure*}

This initial model for the source surface brightness as a function of frequency was then used to calibrate the entire data set, from which a new model using all of the available data was made. However, it was found that the data at the edges of the observed frequency range were not useful due to the sharp increase in the RFI environment at the low-end and the roll over in the HBA bandpass at the high-end. The final calibrated dataset, which has a total bandwidth of 27.5 MHz (141 sub-bands) between 109 and 183 MHz, with a central frequency of 146~MHz, was then imaged using MS-MFS clean to simultaneously obtain the total intensity (Stokes I) and spectral index maps for Cygnus A (see Fig. \ref{fig:maps}). The total intensity map has an rms noise of 43 mJy~beam$^{-1}$ and the full width at half maximum (FWHM) beam size of both maps is $3.8\times2.7$~arcsec at a position angle of $-73$~deg east of north.

In addition, we also made separate maps of the 141 individual sub-bands so that we could test various spectral models for the emission mechanisms at the hotspots. For this, we used the same multi-scale method described above to carry out the deconvolution, which provided a model for the source surface brightness distribution at each sub-band. As we planned to compare the hotspot spectra from LOFAR with those obtained with the VLA at higher frequency, but at a more coarse resolution of 4.5 arcsec \citep{carilli91}, we restored our {\sc clean} models with a circular 4.5 arcsec FWHM beam and measured the peak surface brightness at the position of the hotspots for each of the 141 sub-bands. The spectra of the individual sub-bands are presented in Fig.~\ref{fig:hotspot}, where all 141 spectral data points are plotted. The uncertainties of these data points were determined by adding in quadrature the absolute flux density calibration error of about 2 per cent, which dominates the uncertainty, and the rms noise of the individual maps of each sub-band (typically about 200 mJy~beam$^{-1}$).

We note that the flux-density of the hotspots A and D at 151~MHz, are about 10 per cent lower than those measured by \citet{leahy89} using MERLIN at 151 MHz (after correcting for the different beam sizes). However, \citet{leahy89} comment that they may have over-estimated the flux-density of the hotspots, relative to the extended emission, since they used two point sources to describe the hotspot emission in their starting calibration model. Also, since the LOFAR dataset has a much larger number of antennas relative to the MERLIN dataset, the self-calibration process used for the LOFAR dataset will be more robust.

\section{Results}
\label{results}

The total intensity image of Cygnus A shows the expected combination of compact and extended structure that has been previously observed at low radio frequencies \citep{carilli91,lazio06,leahy89}. For the lobe, which is to the west of the central core, the two main hotspots have been detected, with hotspot A at the edge of the lobe being the brighter of the two, whereas the fainter hotspot B is embedded within the lobe emission. The counter-lobe, which is to the east of the central core, shows a great deal of structure that includes hotspot D and enhanced emission around the edge of the counter-lobe. The back-flow and plume emission between the lobe and counter-lobe have also been clearly detected. 

The spectral index map also shows the same general spectral properties of the source that have been seen either at much lower angular resolution or at higher radio frequencies (e.g. \citealt{kassim93,leahy89}), that is, a steepening of the spectral index from around $\alpha = -0.65\pm0.02$ in the lobe and counter-lobe regions to around $\alpha = -0.95\pm0.14$ at the location of the core. The plume region, which extends south of the core, has the steepest spectral index of any part of the source, with $\alpha = -1.25\pm0.07$. This variation in the spectral properties of the source agrees with what we would expect given previously reported spectral ageing models (e.g. \citealt{carilli91}). There is also evidence for spectral structure within the lobes, particularly within the counter-lobe. However, we defer any detailed discussion of these features to a follow-up paper.

\begin{figure*}
\centering
\includegraphics[width=11cm]{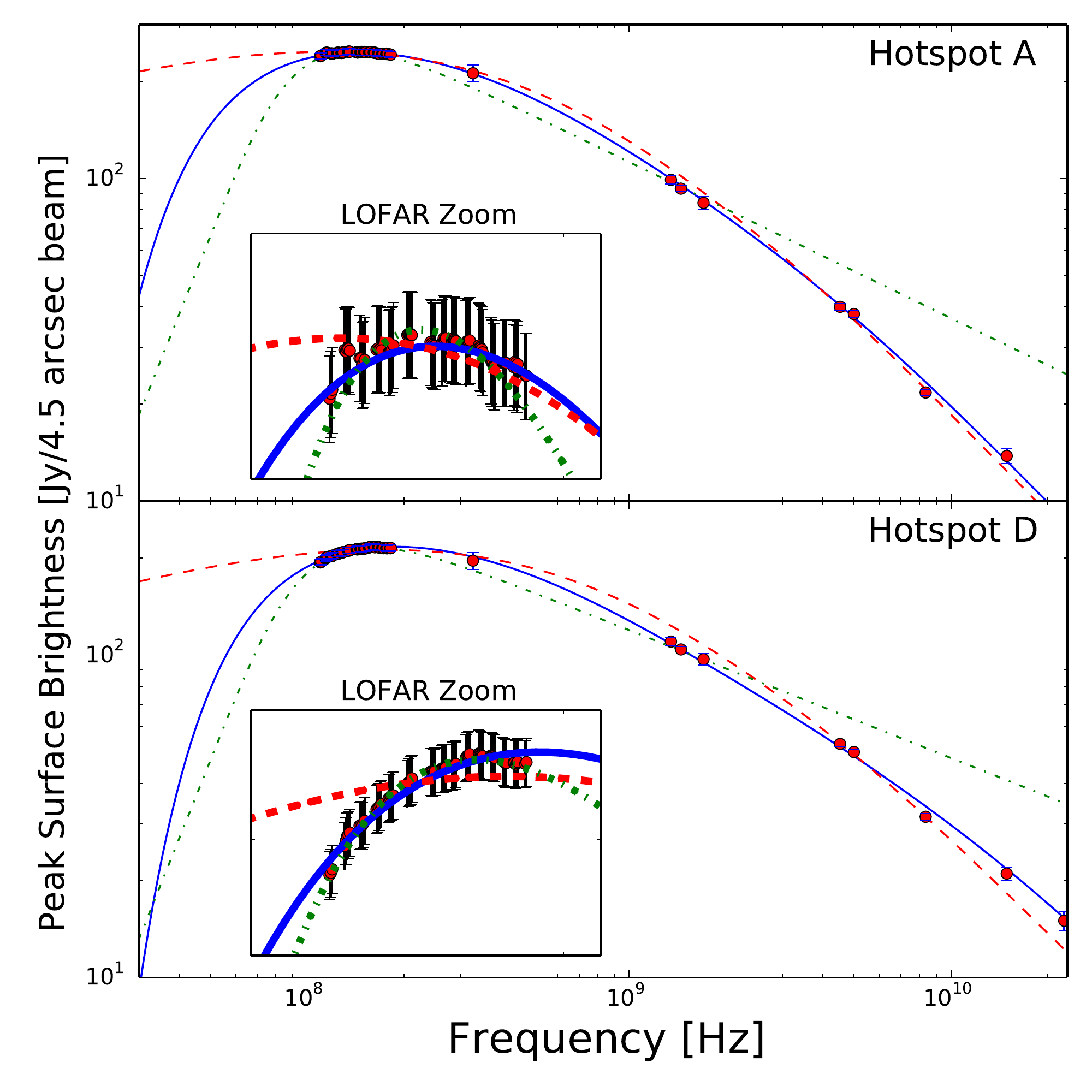}
\caption{The radio spectra of the hotspots A and D of Cygnus A between 109 MHz and 22.45 GHz. The LOFAR data include all 141 spectral points between 109 and 183 MHz grouped in 15 frequency chunks of about 10 sub-bands each, and the higher frequency data are taken from measurements made with the VLA by \citet{carilli91}. For consistency with the VLA measurements, the LOFAR data have been smoothed to a resolution of 4.5 arcsec before the peak surface brightness was measured. Note that this smoothing results in a steepening of the spectral index (with respect to Fig. \ref{fig:maps}) due to the additional contribution from the lobe and counter-lobe emission. Also shown are the fits using a low-energy cut-off model (dashed; red), a free-free absorption model (solid; blue) and a synchrotron self-absorption model (dot-dashed; green). Note that the low-energy cut-off + free-free absorption model (see text) has an identical shape to the free-free absorption model.}
\label{fig:hotspot}
\end{figure*}

The main aim of this paper is to investigate the nature of the two brightest hotspots of Cygnus A by using the high angular resolution and large observed bandwidth provided by LOFAR. It is clear from the spectral index map that both of the main hotspots have inverted radio spectra between 109 and 183 MHz. The average spectral index of hotspot D in the LOFAR band is $\alpha = +$0.36\,$\pm$\,0.02 and for hotspot A is $\alpha = +$0.18\,$\pm$\,0.01, as measured at the position of the peak surface brightness emission in the total intensity map for each hotspot. In Fig.~\ref{fig:hotspot}, we show the spectral energy distribution of the hotspots A and D from combining the LOFAR data between 109 and 183 MHz with the measurements made with the VLA between 327.5 MHz and 22.45 GHz by \citet{carilli91}. We see that the spectra of hotspots A and D turnover at around 140 and 160 MHz, respectively. This is the first time that such a turnover has been directly observed for both of the hotspots of Cygnus A. We also note that the average spectral index for hotspot B in the LOFAR band is $\alpha = -$0.26\,$\pm$\,0.01, which is likely a lower limit due to contamination from the steep spectrum lobe emission. The spectral indices of hotspots A, B, and D are all confirmed to be flatter than is predicted from diffusive shock acceleration models \citep{bell78}. These measurements highlight the important role LOFAR can have in measuring the arcsecond-scale spectra of radio-loud active galaxies at low radio frequencies.

\begin{table*}
 \caption{Maximum posterior model parameters for the free-free absorption (FFA), synchrotron self-absorption (SSA) and low-energy cut-off (LEC) models. Note that where appropriate, we fix the volume, magnetic field strength and $\gamma_{\rm max}$ to $V^A = 1.638 \times 10^{59}$~m$^3$, $B^A = 160~\mu$G, $\gamma_{\rm max}^A = 1.3 \times 10^{5}$ and $V^D = 6.30 \times 10^{58}$~m$^3$, $B^D = 250~\mu$G, $\gamma_{\rm max}^D = 1.3 \times 10^{5}$, for hotspot A and D, respectively. The normalization, $S_{\nu_0}$ and the optical depths are stated at 151 MHz.}
 \label{table:model_parameters}
  \begin{tabular}{ccccc}
  \hline
& FFA & LEC & FFA + LEC & SSA  \\
  \hline
 \multicolumn{5}{l}{Hotspot A}\\
   \hline
$a$					& $2.04\pm0.01$	& $2.04\pm0.02$	& $2.02\pm0.02$	&  $1.97\pm0.04$ \\
$K_e$ [m$^{-3}$]		& $3166\pm122$	& $4293\pm530$	& $3351\pm576$	&  \\
$\gamma_{\rm cr}$		& 				& $791\pm15$		& $637\pm27$		&  \\
$\gamma_b$			& $3476\pm147$	& $2064\pm110$	& $2540\pm197$	&  \\
$\alpha_{\rm FFA/SSA}$	& $-0.52\pm0.01$	& $-0.52\pm0.01$	& $-0.51\pm0.01$	& $-0.48\pm0.01$  \\
$\tau_{\rm FFA/SSA}$	& $0.22\pm0.01$	& 				& $0.06\pm0.01$	& $0.27\pm0.01$   \\
$S_{\nu_0}$ [Jy]		&  				&				&				& $1047\pm31 $    \\

  \hline
 \multicolumn{5}{l}{Hotspot D}\\
   \hline
$a$					&  $2.00\pm0.01$	& $1.97\pm0.01$	& $2.05\pm0.02$	& $1.80\pm0.04$  \\
$K_e$  [m$^{-3}$]		&  $3007\pm56$	& $3594\pm464$	& $4988\pm569$	& \\
$\gamma_{\rm cr}$		&  				& $766\pm15$		& $569\pm18$		& \\
$\gamma_b$			& $5844\pm343$	& $2054\pm139$	& $4149\pm306$	& \\
$\alpha_{\rm  FFA/SSA}$	& $-0.50\pm0.01$	& $-0.49\pm0.02$	& $-0.53\pm0.01$	& $-0.40\pm0.01$    \\
$\tau_{\rm FFA/SSA}$	& $0.29\pm0.01$	&				& $0.11\pm0.01$	& $0.34\pm0.01$   \\
$S_{\nu_0}$ [Jy]		&  				&				&				& $745\pm18$   \\
  \hline
 \end{tabular}
\label{tab:results} 
\end{table*}

\section{Nature of the turnover in the hotspot spectra}
\label{discuss}

The turnover that we observe in the spectra of the hotspots of Cygnus A at low radio frequencies can be explained by an absorption process within the hotspots (or along the line-of-sight) and/or a cut-off in the electron energy distribution at low energies. The two most likely absorption processes are free-free absorption and synchrotron-self absorption. However, as has been discussed above, neither of these two mechanisms can account for the flattening of the hotspot radio spectra that has been seen in previous studies without unrealistic parameter values (e.g. \citealt{leahy89,carilli91,lazio06}). Therefore, models requiring a cut-off in the electron energy distribution at low energies have been preferred (e.g. \citealt{carilli91}).

Using the new improved data provided by LOFAR, we now re-test these interpretations for the low-frequency spectra of hotspots A and D. The model parameters were estimated using a Bayesian 2-level hierarchical analysis, which allows both the measurement errors and the intrinsic scatter in the regression relationship to be correctly accounted for (see \citealt{kelly07} for details). The parameters for the model fits discussed below are presented in Table \ref{tab:results} and the model spectra are shown in Fig.~\ref{fig:hotspot}. In Table~\ref{tab:results}, we also give the relative Bayesian evidence for each model, except for the synchrotron-self absorption model (see below).

\subsection{A low-energy cut-off (LEC) model}

\citet{carilli91} argued that the most likely explanation for the turnover in the hotspot spectra below 1~GHz was a LEC in the electron energy distribution. A LEC may arise due to the dissipation of jet bulk kinetic energy in the hotspots \citep{godfrey09}. To test this interpretation, we model the spectrum by incorporating a LEC in a broken power-law electron spectrum that is inferred from the higher frequency data (e.g. \citealt{carilli91, carilli99}). To model the LEC, we assume a physically realistic low energy tail, rather than an instant cutoff in the electron energy distribution below some cutoff energy, which would produce a more sharp turnover. We use the standard continuous injection model, combined with a low-energy tail, defined as
\begin{eqnarray} \label{eqn:N_gamma}
N(\gamma) &=& N(\gamma_{\rm cr}) \left( \frac{\gamma}{\gamma_{\rm cr}} \right)^2  \qquad  \gamma < \gamma_{\rm cr}  \nonumber \\
N(\gamma) &=& \frac{K_e \gamma_{\rm b}}{(a-1)} \gamma^{-(a+1)} \left[  1 - \left( 1 -  \frac{\gamma}{\gamma_{\rm b}} \right)^{a - 1}  \right]  \qquad  \gamma_{\rm cr} \leq \gamma < \gamma_{\rm b}  \nonumber \\
N(\gamma) &=& \frac{K_e \gamma_{\rm b}}{(a-1)} \gamma^{-(a+1)}   \qquad  \gamma_{\rm b} \leq \gamma < \gamma_{\rm 2}  
\end{eqnarray}
where $N(\gamma)$ is the particle number per unit Lorentz factor, $\gamma$ is the Lorentz factor, $\gamma_{\rm cr}$ is the critical Lorentz factor defining the position of the low energy cutoff, $\gamma_{\rm b}$ is the Lorentz factor at the break frequency of the broken power-law that arises due to the radiative cooling losses, and $a$ is the injection index. 

An immediate cut-off in the electron spectrum below some cutoff energy is unphysical. We have instead assumed a distribution of the form $N(\gamma) \propto \gamma^2$ for electrons below the critical Lorentz factor $\gamma_{\rm cr}$. This low-energy distribution corresponds to the low-energy tail of a relativistic Maxwellian, which is likely to arise if the low energy cut-off is due to the dissipation/thermalisation of jet bulk kinetic energy in the hotspots, as argued by \citet{godfrey09}. This model will produce a more gradual turnover in the radio spectrum than that produced with the classical LEC model which assumes an immediate cutoff in the electron spectrum. We note that a non-uniform magnetic field strength and/or variation in the cutoff energy as a function of position would act to make the turnover even more gradual than our model, and in that sense our model is conservative. The magnetic field strengths for hotspots A and D are assumed to be 160 and 250~$\mu$G, respectively, as given by the synchrotron self-Compton modelling of the hotspot X-ray emission \citep{harris94}.

Given that the other model parameters (injection index and break frequency) are well constrained by the higher frequency data, the shape of the spectrum in the LOFAR dataset is almost entirely determined by the cut-off energy ($\gamma_{\rm cr}$). The best fit model has a $\gamma_{\rm cr}^A = 791\pm15$ and $\gamma_{\rm cr}^D =766\pm15$ for hotspots A and D, respectively.  However, as can be seen from Fig.~\ref{fig:hotspot} (red dashed lines), in striving to fit the LOFAR data, the LEC model performs poorly at higher frequencies, and furthermore, the gradual turnover in the radio spectra produced by the LEC model does not provide a good description of the LOFAR data. Therefore, a LEC model alone, with the chosen shape of the cut-off, cannot be responsible for the observed flattening of the hotspot spectra. The rapid curvature of the observed radio spectrum, and the highly inverted spectrum towards the low-frequency end of the LOFAR data (see Fig.~\ref{fig:maps}) indicates that some form of absorption must be at least partially responsible for the observed turnover.

\subsection{A free-free absorption model} \label{sec:free_free_only}

The rapid downturn in the radio spectra could be related to thermal absorption due to an ionised medium along the line-of-sight to the hotspots, which we now investigate using the following free-free absorption model (e.g. \citealt{kassim89}),
\begin{equation}
S_{\nu} = S_{\nu_0} (\nu/\nu_0)^{\alpha_{\rm FFA}} \exp \left[ -\tau_{\rm \nu_0}\left( \nu / \nu_0 \right)^{-2.1} \right],
\end{equation}
where $S_{\rm \nu_0}$ and $\tau_{\rm \nu_0}$ are the flux density and optical depth at the reference frequency $\nu_0 = 151$~MHz, respectively, and $\alpha_{\rm FFA}$ is the optically thin spectral index. We find the free-free absorption model with radiative cooling to be an excellent fit to the data (see Fig.~\ref{fig:hotspot}), with optical depths at 151 MHz and optically thin spectral indices of $\tau_{\rm FFA}^{A} = 0.22\pm0.01$ and $\alpha_{\rm FFA}^{A } = -0.52\pm0.01$ for hotspot A, and $\tau_{\rm FFA}^{D} = 0.29\pm0.01$ and $\alpha_{\rm FFA}^{D} = -0.50\pm0.01$ for hotspot D, respectively. These optical depths would require particle densities of about $n_e = 3$~cm$^{-3}$ for absorbing clouds with electron temperatures of $8000$~K and sizes of $1$~kpc, which is around an order of magnitude higher than has been observed for the warm intergalactic medium of our own Galaxy (e.g. \citealt{kassim89}). Furthermore, from Fig.~\ref{fig:maps} we see that only the hotspots show inverted radio spectra at frequencies above 100 MHz, which suggests that the absorbing material would have to be localized to those regions only, which seems unlikely if the absorbing material was within our Galaxy. 

Alternatively, the thermal absorbing material may be internal to the hotspots. In this case, assuming an electron temperature of 8000~K and a hotspot diameter of $2$~kpc \citep{wright04}, a particle density of about $n_e = 2$~cm$^{-3}$ would be required. Such high-density thermal material in the hotspots would imply high internal rotation measures. Indeed, \citet{dreher87} obtained an upper limit to the hotspot thermal plasma density of $n_e < 4 \times 10^{-4}$~cm$^{-3}$ based on the low Faraday depolarisation observed at GHz frequencies. They suggest that complex models of magnetic field structure may hide the thermal material to some extent, and increase the limit on the thermal plasma density by up to two orders of magnitude, but even then, the upper limit would still be two orders of magnitude below the density required by the free-free absorption model (see also \citealt{carilli91}). For this reason, the free-free absorption model alone also does not currently provide a realistic explanation to account for the turnover in the radio spectra of the hotspots that is observed with LOFAR.

\subsection{A synchrotron-self absorption model}

The very strong and unresolved emission that is seen from the two hotspots at 146 MHz (see Fig. \ref{fig:maps}) implies brightness temperatures of $\ga 9.5 \times 10^{8}$ and $\ga 8.3 \times 10^{8}$~K for hotspots A and D, respectively, which may suggest that synchrotron-self absorption could be important. We fit the spectra using a synchrotron self absorption model,
\begin{equation}
S_{\nu} = S_{\nu_0} \left( \nu/\nu_0 \right)^{2.5} ( 1 - \exp[-\tau_{\nu_0} \left( \nu / \nu_0 \right)^{\alpha_{\rm SSA} - 2.5}]),
\end{equation}
where $\alpha_{\rm SSA}$ is the optically thin spectral index. The fitted values for the normalization $S_{\nu_0}$ at 151 MHz, injection index $a = 1 - 2 \alpha_{\rm SSA}$ and optical depth $\tau_{SSA}$ at 151 MHz are listed in Table \ref{tab:results}. From this fit, we also obtain the peak frequency ($\nu_p$) and peak flux density ($S_p$). Assuming a single homogenous emitting region of size $\theta = 2$~kpc (e.g. \citealt{wright04}), the inferred magnetic field strengths are calculated using
\begin{equation} \label{eqn:B_SSA}
B_{\rm SSA} \sim (\nu_p/8)^5 S_p^{-2} \theta^4  (1+z)^{-1}
\end{equation}
\citep{kellermann81}, where the peak frequency is measured in GHz. These are found to be $B_{\rm SSA}^A = 0.9$~G and $B_{\rm SSA}^D = 2.6$~G for hotspots A and D, respectively. As has been previously discussed, these magnetic field strengths are $\sim10^4$ times higher than has been estimated from minimum energy arguments (i.e. $\sim300$~$\mu$G; \citealt{carilli91}) or synchrotron-self Compton modelling (i.e. $\sim 200$~$\mu$G; \citealt{harris94}). The synchrotron-self absorption model (green dot-dashed line in Fig.~\ref{fig:hotspot}) does not fit the data at $\ga 2$~GHz, where the observed spectrum steepens, because we do not incorporate a cooling break in the model. This is because with magnetic field strengths of order 1~G, the radiative cooling timescale is prohibitively small, and cannot be the cause of the break in the spectrum observed at GHz frequencies; the break frequency $\nu_b \sim10$~GHz implies a cooling time in the order of months, which is physically impossible given the 2~kpc diameter of the hotspots.  

We note that, as can be seen from equation \ref{eqn:B_SSA}, the derived magnetic field strengths are highly dependent on the assumed peak frequency and size of the emitting region. The peak frequency of the hotspot emission is well-defined from our LOFAR imaging, but the assumption that the emission originates from a single homogenous emitting region of 2~kpc size is known to be incorrect. Observations at 43 GHz with 0.15 arcsec angular resolution find multiple bright components within the hotspots, which also show a non-uniform magnetic field distribution \citep{carilli99}. The angular resolution of the LOFAR imaging presented here is not sufficient to determine the structure of the hotspots at low-frequencies, but we note that if the hotspot emission is dominated by a few components of size 0.2--0.3 kpc, then the derived magnetic fields for a synchrotron self-absorption model could be lowered by several orders of magnitude. However, in order to not overproduce the observed X-ray emission, the individual clumps would need to be highly magnetically dominated, far from minimum energy conditions. Upcoming LOFAR observations including the international stations will achieve 0.2 arcsec resolution, and will provide useful constraints on such a model. 

\subsection{A combination of processes}

A possible explanation for the rapid turnover in the hotspot radio spectra is the combination of a cut-off at low energies and an absorption process. The inclusion of a LEC can reduce the optical depth required, and may bring the derived model parameters into a physically realistic range. In the case of synchrotron-self absorption, the inclusion of a LEC would reduce the effectiveness of the absorption process, because there would be fewer particles to absorb the radiation. Furthermore, the peak flux density and frequency of the emission would not be sufficiently altered to significantly change the magnetic field strength. Therefore, we only consider a combined model that has a LEC with thermal (free-free) absorption. 

We find that such a model provides a good fit to the data (essentially the same as for the free-free model alone; see Fig.~\ref{fig:hotspot}), but with a lower critical Lorentz factor of $\gamma_{\rm cr}^{A} = 637 \pm 27$ and $\gamma_{\rm cr}^{D} = 569 \pm 18$ for hotspots A and D, respectively. As expected, the required optical depths at 151 MHz, and hence particle densities are lower, by a factor of 3 to $\tau_{\rm FFA}^{A} = 0.06\pm0.01$ and $\tau_{\rm FFA}^{D} = 0.11\pm0.01$, corresponding to $n_e = 0.7$~cm$^{-3}$. Although the required optical depths are lower than obtained with the free-free model described in Section \ref{sec:free_free_only}, these values are still several orders of magnitude higher than would be physically realistic if the absorbing material was localized to the region of the hotspots. However, \citet{kassim96} have shown that the integrated spectrum of Cygnus A shows a low frequency turnover at around 20 MHz (see also Fig.~\ref{fig:sed}), which they attribute to free-free absorption by an ionised region within our own Galaxy that covers the extent of the source. They find that the free-free absorption optical depth at 13 MHz is $\tau_{\rm FFA} = 1.2$ for an electron temperature of 10$^4$~K, which is equivalent to $\tau_{\rm FFA} = 0.007$ at 151 MHz. This optical depth is an order of magnitude lower than what we find for the combined free-free absorption model with a LEC.

\section{Conclusions}
\label{conc}

We have presented new low-frequency radio imaging of Cygnus A with LOFAR, which directly shows for the first time, a turnover in the radio spectra of the two secondary hotspots (A and D). We have shown that this turnover cannot result solely from a sharp cut-off in the low-energy distribution of the particles, as was previously suggested, but instead some absorption process must be at least partially responsible. Both synchrotron-self absorption and thermal free-free absorption models provide good fits to the LOFAR data, but both models require implausible physical parameters to produce the observed turnover. Models that combine a low energy cutoff with free-free absorption are also able to provide an excellent fit to the data, but the free-free optical depth remains unrealistically high if the absorbing material is localized to the hotspots. Galactic free-free absorption has an optical depth that is an order of magnitude too low to account for the radio spectrum turnover observed in the LOFAR band. Synchrotron-self absorption in small clumps ($\sim$ 0.2 kpc) may provide an explanation, however such clumps would need to be highly magnetically dominated, far from minimum energy so as to not overproduce the observed X-ray emission from the hotspots. High angular resolution imaging at low radio frequencies ($\la 100$~MHz) with the International LOFAR Telescope (ILT) will enable a strong test of this, and other models for the hotspot spectra in Cygnus A. 

\section*{acknowledgements}
We would like to thank Chris Carilli, Paul Nulsen, {\L}ukasz Stawarz and in particular Dan Harris for useful discussions, and the anonymous referee for their valuable comments. LOFAR, the Low Frequency Array designed and constructed by ASTRON, has facilities in several countries, that are owned by various parties (each with their own funding sources), and that are collectively operated by the International LOFAR Telescope (ILT) foundation under a joint scientific policy. The financial assistance of the South African SKA project (SKA SA) towards this research is hereby acknowledged. Opinions expressed and conclusions arrived at are those of the authors and are not necessarily to be attributed to the SKA SA. The research leading to these results has received funding from the European Research Council under the European Union's Seventh Framework Programme (FP/2007-2013) / ERC Advanced Grant RADIOLIFE-320745.






\textit{\newline
$^{1}$ASTRON, Netherlands Institute for Radio Astronomy, Postbus 2, 7990 AA, Dwingeloo, the Netherlands\\
$^{2}$Kapteyn Astronomical Institute, PO Box 800, 9700 AV Groningen, the Netherlands\\
$^{3}$Max Planck Institute for Astrophysics, Karl Schwarzschild Str. 1, 85741 Garching, Germany\\
$^{4}$Anton Pannekoek Institute for Astronomy, University of Amsterdam, Science Park 904, 1098 XH Amsterdam, the Netherlands\\
$^{5}$School of Physics, Astronomy and Mathematics, University of Hertfordshire, College Lane, Hatfield AL10 9AB\\
$^{6}$University of Hamburg, Gojenbergsweg 112, 21029 Hamburg, Germany\\
$^{7}$Helmholtz-Zentrum Potsdam, DeutschesGeoForschungsZentrum GFZ, Department 1: Geodesy and Remote Sensing, Telegrafenberg, A17, 14473 Potsdam, Germany\\
$^{8}$SRON Netherlands Insitute for Space Research, PO Box 800, 9700 AV Groningen, the Netherlands\\
$^{9}$Max-Planck-Institut f\"{u}r Radioastronomie, Auf dem H\"ugel 69, 53121 Bonn, Germany\\
$^{10}$CSIRO Australia Telescope National Facility, PO Box 76, Epping NSW 1710, Australia\\
$^{11}$Joint Institute for VLBI in Europe, Dwingeloo, Postbus 2, 7990 AA, the Netherlands\\
$^{12}$University of Twente, the Netherlands\\
$^{13}$Harvard-Smithsonian Center for Astrophysics, 60 Garden Street, Cambridge, MA 02138, USA\\
$^{14}$SKA South Africa, 3rd Floor, The Park, Park Road, Pinelands, 7405, South Africa\\
$^{15}$Institute for Astronomy, University of Edinburgh, Royal Observatory of Edinburgh, Blackford Hill, Edinburgh EH9 3HJ\\
$^{16}$Leibniz-Institut f\"{u}r Astrophysik Potsdam (AIP), An der Sternwarte 16, 14482 Potsdam, Germany\\
$^{17}$School of Physics and Astronomy, University of Southampton, Southampton, SO17 1BJ\\
$^{18}$Leiden Observatory, Leiden University, PO Box 9513, 2300 RA Leiden, the Netherlands\\
$^{19}$Hamburger Sternwarte, Gojenbergsweg 112, D-21029 Hamburg, Germany\\
$^{20}$Department of Astrophysics/IMAPP, Radboud University Nijmegen, P.O. Box 9010, 6500 GL Nijmegen, the Netherlands\\
$^{21}$LPC2E - Universite d'Orleans/CNRS\\
$^{22}$Station de Radioastronomie de Nancay, Observatoire de Paris - CNRS/INSU, USR 704 - Univ. Orleans, OSUC , route de Souesmes, 18330 Nancay, France\\
$^{23}$Th\"{u}ringer Landessternwarte, Sternwarte 5, D-07778 Tautenburg, Germany\\
$^{24}$Department of Physics, The George Washington University, 725 21st Street NW, Washington, DC 20052, USA\\
$^{25}$National Radio Astronomy Observatory, 1003 Lopezville Road, Socorro, NM 87801-0387, USA\\
$^{26}$Astronomisches Institut der Ruhr-Universit\"{a}t Bochum, Universitaetsstrasse 150, 44780 Bochum, Germany\\
$^{27}$Astrophysics, University of Oxford, Denys Wilkinson Building, Keble Road, Oxford OX1 3RH\\
$^{28}$Astro Space Center of the Lebedev Physical Institute, Profsoyuznaya str. 84/32, Moscow 117997, Russia\\
$^{29}$National Astronomical Observatory of Japan, Japan\\
$^{30}$Sodankyl\"{a} Geophysical Observatory, University of Oulu, T\"{a}htel\"{a}ntie 62, 99600 Sodankyl\"{a}, Finland\\
$^{31}$STFC Rutherford Appleton Laboratory,  Harwell Science and Innovation Campus,  Didcot  OX11 0QX\\
$^{32}$Department of Physics and Astronomy, University of California Irvine, Irvine, CA 92697, USA\\
$^{33}$Center for Information Technology (CIT), University of Groningen, the Netherlands\\
$^{34}$Centre de Recherche Astrophysique de Lyon, Observatoire de Lyon, 9 av Charles Andr\'{e}, 69561 Saint Genis Laval Cedex, France\\
$^{35}$Jodrell Bank Center for Astrophysics, School of Physics and Astronomy, The University of Manchester, Manchester M13 9PL\\
$^{36}$Department of Physics and Astronomy, University of the Western Cape, Private Bag X17, Bellville 7535, South Africa\\
$^{37}$Department of Physics and Elelctronics, Rhodes University, PO Box 94, Grahamstown 6140, South Africa\\
$^{38}$Department of Astrophysical Sciences, Princeton University, Princeton, NJ 08544, USA\\
$^{39}$LESIA, UMR CNRS 8109, Observatoire de Paris, 92195  Meudon, France
}


\bsp	
\label{lastpage}
\end{document}